\documentclass[dvips,aps,prl,preprint,groupedaddress]{revtex4}
\usepackage{amsmath} 
\usepackage{amsfonts}
\usepackage{amstext}
\usepackage{latexsym, amssymb} 
\usepackage{bm} 
\usepackage[linktocpage=true,colorlinks=true,citecolor=blue]{hyperref} 
\begin{document}
\title{Theory of decoherence due to scattering events and L\'evy
  processes}
\author{Bassano \surname{Vacchini}}
\affiliation{Dipartimento di Fisica dell'Universit\`a di Milano and
  INFN, Sezione di Milano, Via Celoria 16, I--20133, Milan, Italy}
\date{\today}
\begin{abstract}
A general connection between the characteristic function of a L\'evy
process and loss of coherence of the statistical operator
describing the center of mass degrees of freedom of a quantum system
interacting through momentum transfer events with an environment is
established. The relationship with microphysical models and recent
experiments is considered, focusing on the recently
observed transition between a dynamics described by a compound Poisson
process and a Gaussian process.
\end{abstract}
\maketitle
The study of decoherence\cite{KieferNEW}, both at theoretical and
experimental level, owes its relevance to a twofold motivation: on the
one and historically older hand it provides a fruitful research area
for the exploration of the quantum classical boundary, on the other
hand it is the formidable  quantum enemy to be overcome or outwitted
in order to actually realize quantum computers.
Recently various experiments have been performed in which both
qualitative and quantitative analyses 
of decoherence are feasible. These achievements both force and invite
us to go beyond an implicitly established ``common
lore''\cite{ZurekPRA97}, which sometimes deceitfully let features of
simplified models appear universal, contrary to experimental
evidence\cite{RaymerPRL99}. In the present paper we will focus on the
issue of decoherence of the center of mass degrees of freedom of
massive test particle, object of recent and very
accurate quantitative experimental
investigations\cite{MlynekPRL94-PritchardPRL95,PritchardPRL01,ZeilingerQBM-exp,ZeilingerNature04},
showing how these different situations can be addressed within a
unified theoretical approach which, exploring the most often fruitful
connections between quantum and classical
probability\cite{HolevoNEW}, puts into evidence how the loss of
coherence in the off-diagonal position matrix elements of the statistical
operator is generally described by the
characteristic function (CF) of a L\'evy process (LP). The common feature of the
abovementioned experiments is the fact that, provided dissipative
effects which take place on a much longer time scale are neglected,
the interaction causing decoherence can be characterized through momentum
transfer events, which following\cite{AlickiPRA02} we will generally
call collisions; their effect can be described by means of a
decoherence superoperator, a completely positive operation whose
matrix elements in the position representation are given by a quantity
often called decoherence function. In the Markovian case a common
description of such dynamics can be obtained referring to the general
structure of translation-covariant quantum-dynamical semigroups
obtained by Holevo\cite{HolevoRAN-HolevoJMP}, relying on a quantum
generalization of the L\'evy-Khintchine formula. LP are a class of
processes, including Gaussian processes, which despite obeying the
Chapman-Kolmogorov equation characterizing Markov processes not
necessarily have finite variance, so that the central limit theorem
does not always apply. Such processes were in fact found looking for
generalizations of such theorem, and are both space and time
homogeneous, thus naturally arising when considering space translation
invariance. The general structure of the CF, i.e. the Fourier
transform of the probability density (PD), of such processes is given by
the famous L\'evy-Khintchine formula (for a most compact presentation
see\cite{Petruccione} and references therein). The relevance of LP in
physics is growing\cite{ShlesingerLNPH95}, since they allow to cope
with situations not encompassed by the central limit theorem. This is
therefore a natural way to improve the usual, almost ubiquitous models
relying on linear coupling and Gaussian statistics, whose limitations
in the description of open systems and in particular of decoherence
begin to be
appreciated\cite{ZurekPRA97,RaymerPRL99,KusnezovPhysicaE01-BrumerPRL03}.
\par
We first start by introducing in a way adapted to our purposes the
results by Holevo\cite{HolevoRAN-HolevoJMP}, later connecting them to
microphysical derivations and experimental realizations. If the
dynamics causing decoherence is Markovian and described in terms of
momentum transfers, so that in the absence of an external potential one
has translational invariance, the generator of the quantum-dynamical
semigroup generally has the structure ${d\hat\rho}/{dt}=\mathcal{L}_{G}[{\hat \rho}]+\mathcal{L}_{P}[{\hat \rho}]$
with $\hat{\rho}$ the statistical operator of the test particle;
$\mathcal{L}_{G}$ a so-called Gaussian component given by
\begin{equation}
   \label{eq:2}
   \mathcal{L}_{G}[{\hat
     \rho}]=-ia[\hat{x},\hat{\rho}]-\frac{1}{2}D [\hat{x},[\hat{x},\hat{\rho}]],
\end{equation}
written for simplicity in the one dimensional case, with $a\in
\mathbb{R}$, $D>0$ and $\hat{x}$ the position operator of the test
particle; $\mathcal{L}_{P}$ the so-called Poisson component
\begin{multline}
   \label{eq:3}
   \mathcal{L}_{P}[{\hat{\rho}}]=\int dq\, |\lambda (q)|^2 \left[ 
e^{{i\over\hbar} q\hat{x}} \hat{\rho} e^{-{i\over\hbar} q\hat{x}} -\hat{\rho}
\right]
\\
+2\int dq\, \Re (\omega(q) \lambda^*  (q)) \left[ 
e^{{i\over\hbar} q\hat{x}} \hat{\rho} e^{-{i\over\hbar} q\hat{x}} -\hat{\rho}
\right]
\\
+\int dq\, |\omega (q)|^2 \left[ 
e^{{i\over\hbar} q\hat{x}} \hat{\rho} e^{-{i\over\hbar} q\hat{x}} -\hat{\rho}
-\frac{i}{\hbar}\frac{q[\hat{x},\hat{\rho}]}
{1+{q^2}/{q^2_0}}
\right]
\end{multline}
where $|\omega (q)|^2 dq$ is a positive measure, also called L\'evy
measure, with $|\omega (q)|^2$ possibly divergent in zero but such
that the L\'evy condition $\int dq\, |\omega (q)|^2 q^2/ ({1+q^2}) < \infty$
holds, 
the weights $\lambda (q)$ and $\omega (q)$ are in the general case
complex functions, the integration variable $q$ has the dimension of
momentum and the meaning of momentum transfer, the
parameter $q_0$ only appearing for dimensional purposes in the
regularizing factor. In stating the result we have neglected free
evolution and dissipative effects which are relevant only on a much
longer time scale, so that the
momentum of the test particle has essentially been treated as a
$\mathbb{C}$-number. Focusing on the position matrix elements the master-equation takes the form
\begin{equation}
   \label{eq:5}
   \frac{d}{dt}\langle x|\hat{\rho}|y\rangle=\Psi (x-y)\langle x|\hat{\rho}|y\rangle
\end{equation}
with $\Psi (x-y)$ given by (note the dependence on $x-y$ according to
translation invariance)
\begin{multline}
   \label{eq:6}
\Psi (x-y)=-ia (x-y)   -\frac{1}{2}D (x-y)^2
\\
+\int dq\, |\lambda (q)|^2 \left[ 
e^{{i\over\hbar} q(x-y)} -1\right]
\\
+2\int dq\, \Re (\omega (q) \lambda^* (q))\left[ 
e^{{i\over\hbar} q(x-y)} -1\right]
\\
+\int dq\, |\omega (q)|^2\left[ 
e^{{i\over\hbar} q(x-y)} -1-\frac{i}{\hbar}\frac{q (x-y)}
{1+{q^2}/{q^2_0}}
\right],
\end{multline}
so that one immediately has the general solution
\begin{equation}
   \label{eq:7}
   \langle x|\hat{\rho}_t|y\rangle=e^{t\Psi (x-y)}\langle x|\hat{\rho}_0|y\rangle.
\end{equation}
The function $\Phi (t,x-y)\equiv e^{t\Psi (x-y)}$ is the CF of a
LP, $\Psi (x-y)$ being called its characteristic exponent,
the quantity actually fully characterized by the L\'evy-Khintchine
formula. The fact that $\Phi (t,x-y)$ is a CF automatically entails
that its modulus is less than one, the value one for
$x-y$ tending to zero and the value zero for the distance $x-y$
growing to infinity (if a probability density actually exists), i.e. the natural properties in order to predict
the reduction of the off-diagonal matrix elements in~\eqref{eq:7} due
to decoherence. This suppression of coherence however happens with a
variety of behaviors going far beyond the quadratic common lore
corresponding to Gaussian statistics, depending on the process
characterizing the physical interaction.
\par
We now briefly present some microphysical models giving specific
realizations of~\eqref{eq:5} and make later contact with actual experiments; as it turns
out Eq.\eqref{eq:5} actually encompasses all known models of decoherence
for the center of mass degrees of freedom\cite{KieferNEW}. Let us
first consider the motion of a massive test particle interacting
through collisions with a background gas, developed in detail
in\cite{art3-art5-art10}, where also dissipative effects have been
taken into account, relying on a kinetic approach. Neglecting free
motion and dissipation the result becomes
\begin{equation}
   \label{eq:8}
 \frac{d}{dt}\langle \bm{x}|\hat{\rho}|\bm{y}\rangle=
 n (2\pi)^4 \hbar^2  
   \int d^3\!
        \bm{q}
        \,  
        {
        | \tilde{t} (q) |^2
        }
        S(\bm{q},E)
\left[ 
e^{{i\over\hbar}\bm{q}\cdot (\bm{x}-\bm{y})}-1
\right]\langle \bm{x}|\hat{\rho}|\bm{y}\rangle
\end{equation}
where $n$ is the gas density, $\tilde{t} (q)$ the Fourier transform of
the interaction potential and $S$ a two-point correlation function
characterizing the gas known as dynamic structure factor depending on
both momentum and energy transfer ($\bm{q}$ and $E$). For a finite
macroscopic scattering cross-section $\sigma= (2\pi)^4 \hbar^2 ({M}/{p_0}) \int d^3\!  \bm{q} \,
{ | \tilde{t} (q) |^2 } S(\bm{q},E)$,
with $M$ the mass of the test particle and ${p_0}$ its incoming
momentum, one can introduce a scattering rate $\Lambda\equiv n
({p_0}/M) \sigma$
and a suitably normalized PD
\begin{equation}
   \label{eq:11}
   \mathcal{P} (\bm{q})=\frac{n}{\Lambda}  (2\pi)^4 \hbar^2
        {
        | \tilde{t} (q) |^2
        }
        S(\bm{q},E),
\end{equation}
so that~\eqref{eq:7} reads
\begin{equation}
   \label{eq:13}
   \langle \bm{x}|\hat{\rho}_t|\bm{y}\rangle=e^{\Lambda t \left[
        \Phi_{\scriptscriptstyle \mathcal{P}}(\bm{x}-\bm{y})   
  -1\right]}\langle \bm{x}|\hat{\rho}_0|\bm{y}\rangle,
\end{equation}
where we have introduced the CF $\Phi_{\scriptscriptstyle \mathcal{P}}$
associated to the PD $\mathcal{P}$, i.e. its Fourier transform.  Here no
confusion should arise: the exponential function in~\eqref{eq:13} is
the CF of a LP which in this particular case can be expressed in terms
of the CF $\Phi_{\scriptscriptstyle \mathcal{P}}$ of the PD $\mathcal{P}$.
Eq.\eqref{eq:13} is a particular realization of~\eqref{eq:7} given by
the choice $a=D=0$, $\omega (q)=0$ and $|\lambda (q)|^2\rightarrow\Lambda
\mathcal{P}  (\bm{q})$ in~\eqref{eq:6}, corresponding to a compound
Poisson process\cite{FellerII}.  The physical picture behind it is
the following: the dynamics is driven by collisions, the probability
of having a definite number of collisions in a time $t$ being given by
a Poisson distribution with intensity $\Lambda$ and mean $\Lambda t$,
each collision however is not characterized by a fixed, deterministic
value of the transferred momentum $\bm{q}$, but rather by a certain PD
$\mathcal{P}  (\bm{q})$ depending in the case under consideration on the two-body
interaction potential and a suitable correlation function.
Leaving aside for a moment the detailed structure of~\eqref{eq:8}
related to its microphysical derivation, the result~\eqref{eq:13}
generally applies to a situation in which one has a collection of
momentum transfer events each characterized by a certain PD (to be
obtained or introduced by means of some microscopic or
phenomenological model) corresponding to a compound Poisson process.
Note that the fact that the probability of having a certain number of
events is Poisson distributed is crucial in order to have a Markovian
dynamics\cite{HerzogPRA95}, as we shall see later on.  The
result~\eqref{eq:13} embraces the work by Gallis and
Fleming\cite{GallisPRA90}, which apart from a simple but relevant
correction\cite{ZeilingerQBM-th-garda03} has been used for the
theoretical analysis of decoherence experiments with fullerenes, both
in the case of collisional decoherence\cite{ZeilingerQBM-exp} and of
decoherence due to thermal emission of
radiation\cite{ZeilingerNature04}. Both situations correspond to
compound Poisson processes, where the relevant PD $\mathcal{P}  (\bm{q})$ is
obtained in terms of the collisional cross-section and the spectral
photon emission rate respectively\cite{commento-GRW}. According to a detailed theoretical
analysis\cite{ZeilingerQBM-th2} the final visibility is obtained by
an average of the characteristic exponent in~\eqref{eq:13} over the
possible path separations in the interferometer. Furthermore in the
case of collisional decoherence the random momentum kicks are so
strong that the CF $\Phi_{\scriptscriptstyle \mathcal{P}}$ in~\eqref{eq:13} is
essentially zero for the path separations of interest, so that its
actual structure is not relevant and only the mean $\Lambda t$
determines the fringes visibility.
The connection of Eq.\eqref{eq:13} with the
common lore of a Gaussian process is
straightforward\cite{art3-art5-art10}, expanding the exponential
in~\eqref{eq:8} up to second order
the solution rather than~\eqref{eq:13} becomes
\begin{equation}
   \label{eq:15}
   \langle \bm{x}|\hat{\rho}_t|\bm{y}\rangle=
e^{\Lambda t \left[
   i \langle\bm{q}\rangle\cdot(\bm{x}-\bm{y})  
-\frac{1}{2} 
\sum
\langle q_i q_j\rangle (x_i-y_i) (x_j-y_j)\right]}\langle \bm{x}|\hat{\rho}_0|\bm{y}\rangle,
\end{equation}
where $\langle\bm{q}\rangle\equiv         \int d^3\!
        \bm{q}
        \,  \mathcal{P}  (\bm{q}) \bm{q}$ and $\langle q_i q_j\rangle 
\equiv         \int d^3\!
        \bm{q}
        \,  \mathcal{P}  (\bm{q}) q_i q_j$
are the moments of the PD $\mathcal{P}$ appearing by
definition (if they exist) as coefficients in the Taylor expansion of
the CF $\Phi_{\scriptscriptstyle \mathcal{P}}$. One thus ends up 
with the CF of a Gaussian process
with mean given by the product of the intensity $\Lambda$ and the
first moment of the distribution $\mathcal{P}$ characterizing the original
compound Poisson process, and variance given by the product of
intensity and second moments, corresponding to the choice
$a\rightarrow -\Lambda  \langle\bm{q}\rangle$ and $D\rightarrow
D_{ij}=\Lambda \langle q_i q_j\rangle$ in \eqref{eq:6}, 
$\lambda (q)$ and $\omega (q)$ being zero. As a
last example we consider the case of a massive test particle
interacting with a chaotic environment, modelled through random
matrices. In the absence of an external potential and considering an
environment with constant average level density the dynamics is given
by\cite{KusnezovPRL99}
\begin{equation}
   \label{eq:17}
   \frac{d}{dt}\langle x|\hat{\rho}|y\rangle=K  \left[G \left(\frac{x-y}{x_0} \right)-1\right]\langle x|\hat{\rho}|y\rangle
\end{equation}
where $G$ is directly related to a two-point correlation
function describing the chaotic background, with characteristic
correlation length $x_0$, while $K$ is a coupling constant. In the
weak-coupling limit the authors of\cite{KusnezovPRL99} propose the
expression $G (r)\approx 1- |r|^{\alpha}$
requiring $\alpha\in (0,2]$ due to some necessary restriction on the
two-point correlation function, so
that~\eqref{eq:17} has the simple solution
\begin{equation}
   \label{eq:19}
   \langle x|\hat{\rho}_t|y\rangle=
e^{-Kt\left|\frac{x-y}{x_0}\right|^\alpha}
\langle x|\hat{\rho}_0|y\rangle.
\end{equation}
They then show that in this case the statistical operator can display
dynamics given by so-called L\'evy stable laws. This is a naturally
expected result in the present framework since \eqref{eq:19} is a
particular case of~\eqref{eq:7} corresponding to $a=D=0$, $\lambda
(q)=0$ and a  L\'evy measure 
$|\omega (q)|^2 \propto 1/ |q|^{\alpha+1}$
corresponding to the symmetric stable LP with
scaling exponent $\alpha$\cite{Bertoin}; the
restriction on $\alpha$ now arises due to the  L\'evy
condition, the case $\alpha=2$ corresponding to a Gaussian process,
all other symmetric L\'evy stable laws having infinite second moments.
Suppression of spatial coherence for random momentum transfers
governed by a L\'evy stable law is expected to be stronger than for
the usual Gaussian case\cite{LutzPLA02}, even though no experimental
evidence is available yet.
\par
We now consider
the transition between~\eqref{eq:13} and~\eqref{eq:15}, i.e. from the
CF of a compound Poisson process to that of the
related Gaussian process, in view of recent experiments on decoherence
in an atom interferometer, obtained by spontaneous scattering of photons
off atoms interacting in a controlled way with a
laser\cite{MlynekPRL94-PritchardPRL95}. We will focus in particular
on the most recent results\cite{PritchardPRL01} in which both single- and multiple-photon
decoherence has been observed, noting that ``the few-photon limit is
of a qualitatively different character'' and following ``the smooth
transition between these two regimes'', connecting the many photon
limit with the ``common lore'' master-equation\cite{RaymerPRL99}
predicting exponential reduction in coherence with separation
squared. For the case of an atom interacting with a laser the
PD that the atom experiences a given momentum
transfer along the direction of propagation of the laser as a
consequence of spontaneous emission has been characterized by
Mandel\cite{MandelJO79}, let us call it $\mathcal{P}_{\scriptscriptstyle M} (q)$ for the case of a
single photon. We can now therefore write the master-equation for the
case at hand in analogy to~\eqref{eq:8} in operator form as follows
\begin{equation}
   \label{eq:21}
   \frac{d}{dt}\hat{\rho} =\Gamma \int dq\, \mathcal{P}_{\scriptscriptstyle M} (q)\left[ 
e^{{i\over\hbar} q\hat{x}} \hat{\rho} e^{-{i\over\hbar} q\hat{x}} -\hat{\rho}
\right],
\end{equation}
where $\Gamma$ is once again a scattering rate depending e.g. on the
intensity of the laser. In order to make contact with the analysis put
forward in\cite{PritchardPRL01} we formally write the solution as a Dyson
expansion, thus describing the time evolution as a sequence of jumps,
given by the random momentum transfers described by $\mathcal{P}_{\scriptscriptstyle M} (q)$, on the
background of a relaxing evolution, trivial for the case at hand in
which we neglect free dynamics and dissipation.  The jump expansion
reads
\begin{multline}
   \label{eq:22}
   \hat{\rho}_t=e^{-\Gamma t}\hat{\rho}_0 
+ \sum_{n=1}^{\infty}
\int_0^{t} dt_{n}\int_0^{t_{n}} dt_{n-1} \ldots \int_0^{t_{2}} dt_{1}
\\
\times
e^{-\Gamma (t-t_{n})}\Gamma \mathcal{J}_{\mathcal{P}_{\scriptscriptstyle M}}e^{-\Gamma
  (t_{n}-t_{n-1})}\ldots e^{-\Gamma (t_{2}-t_{1})}\Gamma
\mathcal{J}_{\mathcal{P}_{\scriptscriptstyle M}}e^{-\Gamma t_{1}}\hat{\rho}_0
\\
=\sum_{n=0}^{\infty} \frac{(\Gamma t)^n}{n!}e^{-\Gamma t}
\underbrace{\mathcal{J}_{\mathcal{P}_{\scriptscriptstyle M}} \circ \ldots \circ
  \mathcal{J}_{\mathcal{P}_{\scriptscriptstyle M}}}_{n\ \mathrm{times}}[\hat{\rho}_0],
\end{multline}
with $\mathcal{J}_{\mathcal{P}_{\scriptscriptstyle M}}$ a decoherence superoperator given by the
following completely positive, trace preserving operation
\begin{equation}
   \label{eq:23}
   \mathcal{J}_{\scriptscriptstyle \mathcal{P}_{\scriptscriptstyle M}}[\hat{\rho}]\equiv \int dq\, \mathcal{P}_{\scriptscriptstyle M} (q)
e^{{i\over\hbar} q\hat{x}} \hat{\rho} e^{-{i\over\hbar} q\hat{x}},
\end{equation}
where $\mathcal{P}_{\scriptscriptstyle M}$ is a PD and the $e^{{i\over\hbar}
  q\hat{x}}$ are momentum translation operators. This decoherence
superoperator generally describes the effect on the statistical
operator of a momentum transfer randomly distributed according to
$\mathcal{P}_{\scriptscriptstyle M}$. The matrix elements of the decoherence superoperator in the
position representation give a function often called decoherence
function\cite{ZeilingerQBM-exp,PritchardPRL01}, actually the
CF associated to $\mathcal{P}_{\scriptscriptstyle M}$, with all its natural
properties, including the fact that it is positive definite,
corresponding to the complete positivity of the decoherence
superoperator $\mathcal{J}_{\scriptscriptstyle \mathcal{P}_{\scriptscriptstyle M}}$. 
The Mandel PD $\mathcal{P}_{\scriptscriptstyle M}$
leads to
\begin{displaymath}
\Phi_{\mathcal{P}_{\scriptscriptstyle M}} (x-y)
=\frac{3}{2}e^{ik_0 (x-y)}
\left\{
\mathop{\mathrm{sinc}}\nolimits[k_0 (x-y)] + \frac{\cos[k_0
  (x-y)]-\mathop{\mathrm{sinc}}\nolimits[k_0 (x-y)]}{[k_0 (x-y)]^2} 
\right\},
\end{displaymath}
with $k_0$ the wave vector of the exciting light, and using
\begin{displaymath}
   \underbrace{\mathcal{J}_{\mathcal{P}_{\scriptscriptstyle M}} \circ \ldots \circ
  \mathcal{J}_{\mathcal{P}_{\scriptscriptstyle M}}}_{n\ \mathrm{times}}[\hat{\rho}]
=
\int dq\, ( \underbrace{\mathcal{P}_{\scriptscriptstyle M} \ast \ldots \ast
  \mathcal{P}_{\scriptscriptstyle M}}_{n\ \mathrm{times}}) (q)e^{{i\over\hbar} q\hat{x}} \hat{\rho} e^{-{i\over\hbar} q\hat{x}},
\end{displaymath}
whit $\circ$ the composition of superoperators and $\ast$
the convolution of PD (the convolution $n$ times of
$\mathcal{P}_{\scriptscriptstyle M}$ giving according to\cite{MandelJO79} the PD
that a momentum transfer $\bm{q}$ is imparted to the atom as a
consequence of $n$ spontaneous emissions), the matrix elements of
Eq.\eqref{eq:22} become
\begin{align}
   \label{eq:27}
   \langle x|\hat{\rho}_t|y\rangle &=\sum_{n=0}^{\infty} \frac{(\Gamma
     t)^n}{n!}e^{-\Gamma t}\Phi_{\mathcal{P}_{\scriptscriptstyle M}}^n (x-y)\langle
   x|\hat{\rho}_0|y\rangle
\\
\nonumber
&\equiv \sum_{n=0}^{\infty} p_n (t) \Phi_{\mathcal{P}_{\scriptscriptstyle M}}^n (x-y)\langle
   x|\hat{\rho}_0|y\rangle
\end{align}
where according to the property of the Fourier transform the $n$-th
power of the CF $\Phi_{\mathcal{P}_{\scriptscriptstyle M}}$ appears; note that if the scattering
rate is assumed time dependent, according to a time inhomogeneous
Poisson process, nothing would change but the replacement $\Gamma
\rightarrow \int_{0}^{t} dt' \Gamma (t')$.  If the $p_n (t)$ are Poisson
distributed with mean $\bar{n}\equiv \Gamma t$, where $t$ is the time
of interaction with the laser, Eq.\eqref{eq:27} is exactly equivalent
to Eq.\eqref{eq:13} and this is the only distribution of the weights
$p_n (t)$ describing a Markovian dynamics\cite{HerzogPRA95}. For the decoherence experiments
in atom interferometry\cite{PritchardPRL01} the relative contrast is
directly related to the modulus of the CF in~\eqref{eq:7}, so that
provided the dynamics is Markovian switching from the single- to the
many-photon limit for growing intensity of the laser the compound
Poisson process characterized by $\Phi_{\mathcal{P}_{\scriptscriptstyle M}}$, and described
by~\eqref{eq:13} or~\eqref{eq:27}, goes over to the related Gaussian
process described by~\eqref{eq:15}. This is essentially what has been
observed for the first time in\cite{PritchardPRL01}: the smooth
transition between the two qualitatively distinct regimes can
therefore be understood and described in a unified way on the basis of the presented
theoretical framework, expressing the loss of spatial coherence in
terms of the CF of a suitable LP. In particular the authors
of\cite{PritchardPRL01} compare their results with the
master-equation only in the many-photon limit, when the ``common
lore'' quadratic expression applies, apart from the correction due to
the non vanishing first moment 
reflecting anisotropy; in the single- or few-photon limit they rely on
a formula like the jump expansion~\eqref{eq:27} of the
master-equation, fitting from the very beginning the experiment with a
Gaussian distribution for $p_n (t)$ (though possibly allowing for a
Poisson relationship between mean and variance), rather than with a
Poisson distribution corresponding to the ideal
case\cite{Walls-Milburn} which describes a Markovian dynamics. These
small corrections notwithstanding, depending on deviations of the atom
laser interaction from the Markov regime, these experiments have obtained
the first experimental study of the transition between the decoherence
regimes described by Eq.\eqref{eq:13} and Eq.\eqref{eq:15}
respectively.
\par
A general theoretical description of
decoherence due to random momentum transfers has been presented, showing how spatial
coherence is suppressed according to the CF of a
LP. This has been obtained relying on the general
structure of translation-covariant generators of quantum-dynamical
semigroups derived by Holevo as a quantum L\'evy-Khintchine formula. Different microscopic
models have been shown to lead to particular examples
of the general structure; not only Gaussian processes, but also
compound Poisson and symmetric stable LP have been
considered, thus going beyond the usual limitation given by Gaussian
statistics and opening the way for both microscopical and
phenomenological treatment of new scenarios, especially in connection
with chaotic environments. 
%
%
\begin{acknowledgments}
I am really indebted to Prof. L. Lanz and  Prof. A.  Barchielli for very useful
   discussions on the subject of this paper.  This work was
   financially supported by MIUR under Cofinanziamento and FIRB.
\end{acknowledgments}

\begin{thebibliography}{99}
\bibitem{commento-GRW} Note that despite appearances and the
  completely different meaning the famous GRW master-equation is also
  of this kind, with the most natural choice: the distribution $P
  (\bm{q})$ of random momentum transfers is Gaussian.
\end{thebibliography}

\end{document}